\documentstyle[11pt,twocolumn,psfig]{nature}

\def\be{\begin{equation}}
\def\ee{\end{equation}}

\def\msun{{M_\odot}}

\def\gsim{\lower.5ex\hbox{\gtsima}}
\def\lsim{\lower.5ex\hbox{\ltsima}}
\def\gtsima{$\; \buildrel > \over \sim \;$}
\def\ltsima{$\; \buildrel < \over \sim \;$}
\def\prosima{$\; \buildrel \propto \over \sim \;$}
\def\gsim{\lower.5ex\hbox{\gtsima}}
\def\lsim{\lower.5ex\hbox{\ltsima}}
\def\simgt{\lower.5ex\hbox{\gtsima}}
\def\simlt{\lower.5ex\hbox{\ltsima}}
\def\simpr{\lower.5ex\hbox{\prosima}}
\def\la{\lsim}
\def\ga{\gsim}

\def\beq#1{\begin{equation}\label{#1}}
\def\eeq{\end{equation}}
\def\beqa#1{\begin{eqnarray}\label{#1}}
\def\eeqa{\end{eqnarray}}

\def\H2p{H$_2^+$ }

\def\mH2p{H_2^+}

\title{LOW-MASS RELICS OF EARLY STAR FORMATION}
\mainauthor{}	

\author{{R. Schneider\affiliation{Osservatorio Astrofisico di Arcetri, Largo Enrico Fermi 5, 
	  50125 Firenze, Italy}\affiliation{``Enrico Fermi'' Centre, Via Panisperna 89/A, 00184 Roma, Italy}, 
A. Ferrara\affiliation{SISSA/International School for Advanced Studies,  
Via Beirut 4, 34100 Trieste, Italy}, R. Salvaterra$^{\ddag}$, K. Omukai\affiliation{Division of Theoretical 
Astrophysics, National Astronomical Observatory, Mitaka, Tokyo 181-8588, Japan} and V. Bromm\affiliation{Harvard-Smithsonian Center for Astrophysics, 60 Garden Street, Cambridge, Massachusetts 02138, USA}}}

\begin{document}

\summary{The earliest stars to form in the Universe were the first sources of light, heat and metals
after the Big Bang. The products of their evolution will have had a profound impact on 
subsequent generations of stars. Recent studies
\cite{Omukai98}$^-$\cite{Omukai03} of primordial star 
formation have shown that, in the absence of metals (elements heavier than helium), 
the formation of  stars with masses 100 times that of the Sun
would have been strongly favoured, and that low-mass stars could not have formed before a
minimum level of metal enrichment had been reached. The value of this minimum
level is very uncertain, but is likely to be
between $10^{-6}$ and $10^{-4}$ that of the Sun\cite{Schneider02}$^,$\cite{Bromm01}.
Here we show that the recent discovery\cite{Christlieb02} of the most iron-poor star known 
indicates the presence of dust in extremely low-metallicity gas, and that this dust is  
crucial for the formation of lower-mass second-generation stars that could survive until today.
The dust provides a pathway for cooling the gas that leads to fragmentation of the precursor
molecular cloud into smaller clumps, which become the lower-mass stars.}
%%%%%%%%%%%
\maketitle
\raisebox{13cm}[-13cm]{\emph{Nature} {\bf 422} (2003), 869--871 (issue 24 April)}
%%%%%%%%%%%
The Hamburg/ESO objective prism survey team has recently reported the discovery\cite{Christlieb02} of a 
star, HE0107-5240, with a mass of $0.8$ solar masses, ($0.8 \msun$) and an iron abundance of 
[Fe/H] = $-5.3 \pm 0.2$ (here we adopt the standard definition [A$_i$/A$_j$]=$\log_{10}(N_i/N_j) - \log_{10}(N_i/N_j)_{\odot}$ where the subscript '$\odot$' refers to
solar values). 
This represents the most iron-deficient star observed to date, and it has been found at a distance of about 
11 kpc from the Sun -- a region inaccessible to previous spectroscopic surveys that were limited to the inner
halo of the Galaxy. \\
Does the existence of this $0.8\msun$ very iron-deficient star suggest that the 
first generation of stars also contained low-mass and long-lived objects? 
The star HE0107-5240 does not show any sign of selective dust depletion that may have altered 
its elemental abundances. Being a low-mass star, elements heavier than Mg can not
be produced by post-formation processes, and their chemical composition reflects the composition 
of the gas cloud out of which HE0107-5240 formed. The authors of ref. 9 thus conclude that the star is 
probably a second-generation object that formed from a gas cloud with metal abundance corresponding 
to [Fe/H]~$=-5.3$ that has been pre-enriched by a supernova from a previous generation star. \\
As the metal composition of supernova ejecta depends on the initial mass of the progenitor star,
the observed abundance pattern of HE0107-5240 may represent an important indication of the nature of
the first stars and of the physical processes leading to low-mass star formation in primordial
gas clouds. \\
Two plausible models need to be considered.
If we assume pre-formation origin for all the heavy elements observed in HE0107-5240,
the gas cloud out of which the star formed would have had to be pre-enriched to a metallicity as high as 
$Z=10^{-2}Z_{\odot}$ (where $Z_{\odot}$ is the solar metallicity). This metallicity is significantly higher 
than the proposed minimum level of 
pre-enrichment $Z_{\rm cr} = 10^{-5 \pm 1} Z_{\odot}$ (ref. 6) above which low-mass stars 
can form. \\
In the second model, suggested in ref. 9, the observed abundances of C, N, Na and Mg 
may be due to post-formation mechanisms, such as mass transfer from a companion star, or
self-enrichment according to recent nucleosynthesis models of stars of mass similar
to that inferred for HE0107-5240 ($\sim 0.8 \msun$) of zero or near-zero initial 
metallicity\cite{Seiss02}. In this ``minimal pre-enrichment model'', only elements heavier than Mg
have a pre-formation origin. Indeed, the authors of ref. 9 argue that the abundance pattern of elements 
heavier than Mg is consistent with the predicted elemental yields of a $20-25 \msun$ star that exploded 
as a Type II supernova, indicating that HE0107-5240 may have formed from a gas cloud which had been 
pre-enriched by such a supernova. 
However, if the first stars had a mass within the range
$200 \msun \leq M \leq 220 \msun$ and exploded as pair-creation supernovae\cite{Heger02},  
the observed abundance pattern of HE0107-5240 for elements heavier than Mg can be equally well reproduced. \\
Thus, with the current data it is not possible to uniquely determine
the mass of the first-generation star(s) that produced the metals locked up
in HE0107-5240. To do this, we need improved upper limits on the abundances 
of Zn (which is predicted to be
about two orders of magnitudes smaller in the ejecta of a $200-220 \msun$ pair-creation supernova 
than for a $22 \msun$ Type II supernova) and heavy
r-process elements in HE0107-5240, as well as in other extremely metal-poor
stars to be found in the future.
The r~process is believed to only
operate in an extremely neutron-rich environment\cite{Truran02}, as is provided during
the creation of a central neutron star in a core-collapse (Type II) supernova.
A pair-creation supernova, on the other hand, is fundamentally different\cite{Heger02}, in
that no remnant is created, and that the necessary condition for the r~process
is not realized. Observing the absence or presence of r-process elements,
therefore, holds the promise of directly probing the nature of the first stars\cite{Qian02}.
%%%%%%%%%%%%%%%%%%%%%%%%%%%%%%%%%%%%%%%%%%
\begin{figure*}[t]
\begin{center}
\centerline{\psfig{figure=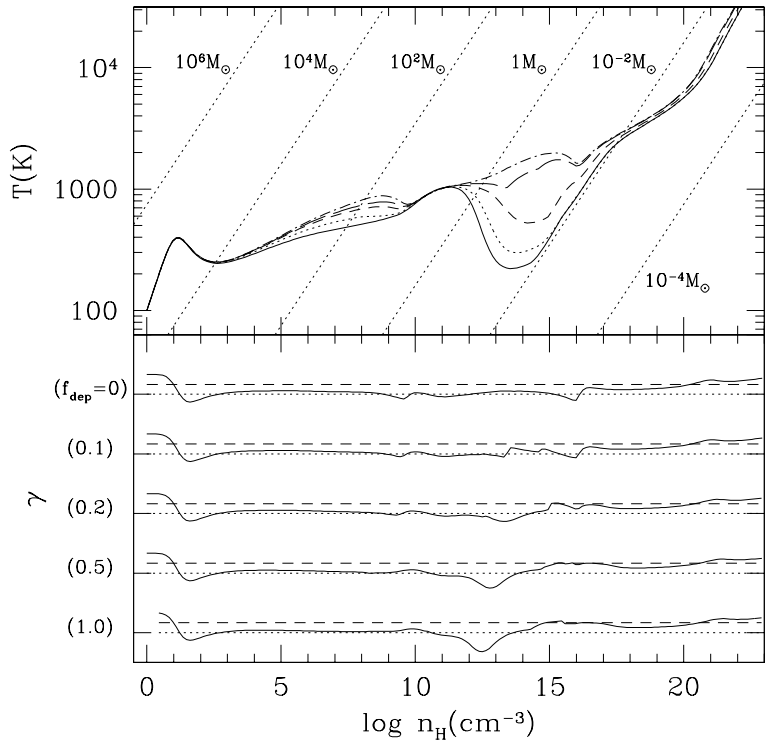,height=11cm}}
\caption{\footnotesize Collapse and fragmentation of low-metallicity star-forming gas clouds. 
 Upper panel, curves show the temperature evolution as a function of the hydrogen number density 
 of proto-stellar clouds with the same initial metallicity $Z=10^{-5.1} Z_{\odot}$, consistent with the
 observed properties of HE0107-5240, but varying dust depletion factors:
 $f_{\rm dep} = 0$ (dash-dotted), $f_{\rm dep} = 0.1$ (long-dashed), 
 $f_{\rm dep} = 0.2$ (short-dashed), $f_{\rm dep} = 0.5$ (dotted) and  $f_{\rm dep} = 1$ (solid). 
 The diagonal dotted lines correspond to constant Jeans mass for spherical clumps. 
 Lower panel,  the adiabatic index $\gamma$ as a function of the hydrogen number density for the curves 
 shown in the upper panel. Dotted (dashed) lines correspond to $\gamma = 1$
 ($\gamma = 4/3$) (see text). Curves are labelled with the appropriate value of $f_{\rm dep}$.}
\label{fig:1}
\end{center}
\end{figure*}
%%%%%%%%%%%%%%%%%%%%%%%%%%%%%%%%%%%%%%
Thus we can ask: what was the actual level of metallicity of the parent cloud out 
of which HE0107-5240 was born?
Assuming the same interstellar medium dilution factor for all the observed heavy elements, we can directly derive 
a lower limit for the metallicity of the cloud.
According to such estimates, the explosions of stars with mass in the ranges 
$22 \msun$ and $200 \msun < M_{\star} <220 \msun$ were 
able to enrich the star-forming gas up to a metallicity $10^{-5.5} \la Z/Z_{\odot} \la 10^{-5.1}$.
This range of initial metallicities falls within the proposed
minimum level of metal pre-enrichment\cite{Schneider02} above which low-mass 
objects can form. Thus, the observation of HE0107-5240 appears to be extremely helpful in determining 
the critical conditions that finally enable low-mass star 
formation at early times. 
Following the analysis of ref. 6, we recall that if some efficient cooling mechanism
is present, the energy deposited by gravitational contraction cannot balance the radiative losses,  
and the gas cloud cools and fragments into increasingly smaller gas clumps. The necessary conditions 
to stop fragmentation and start gravitational contraction within each clump are that cooling becomes 
inefficient, and the Jeans mass of the fragments does not decrease any further, 
thus favouring fragmentation into sub-clumps. 
This  condition depends somewhat on the geometry of the fragments, and translates into
$\gamma >4/3$ ($\gamma > 1$) for spherical (filamentary) clumps, where $\gamma$ is the adiabatic index defined
as $T \propto n^{\gamma-1}$. \\
For a metal-free gas, the only efficient coolant is molecular hydrogen. Cooling due to molecular line emission
becomes inefficient at densities above $n > 10^3 \,\mbox{cm}^{-3}$, and fragmentation stops when the minimum
fragment mass is of order $10^3 -10^{4}\msun$. As the gas becomes more and more enriched with heavy elements, 
the cooling rate at low density $n \leq 10^3 \,\mbox{cm}^{-3}$ increases because of metal (especially O and C) 
line emission. 
More importantly, however, the depletion factor, defined as the dust-to-metals mass ratio 
$f_{\rm dep} = M_{\rm gr}/M_{Z}$, is responsible for activating (through dust-gas thermal exchanges) a 
new strong phase of fragmentation at high density, which finally leads to fragments of low mass. 
For initial metallicities in the range $10^{-4} - 1 Z_{\odot}$, the presence of dust becomes almost irrelevant,
and the minimum fragment masses are in the range $10^{-2} - 1 M_{\odot}$, several orders of magnitude smaller than in the 
metal-free case. \\
In Fig.~1, we show the evolution of a star forming gas cloud with initial metallicity of $Z=10^{-5.1} Z_{\odot}$
as if it were pre-enriched by the pair-creation supernova explosion of a  205 $\msun$ star. 
The various curves correspond to different values for the depletion factor.
The thermal evolution of the gas is insensitive to the presence of different amounts of dust grains 
as long as the density is below $10^5 \,\mbox{cm}^{-3}$. In this regime, the cooling rate is dominated by
molecular line cooling. At higher densities, gas clouds with depletion factors $\ga 0.2$ experience a 
new phase of fragmentation (shown as the dip in the $\gamma$ evolution at $n \approx 10^{12} \,\mbox{cm}^{-3}$), 
which is due to dust-gas thermal exchanges and lasts for about three orders of magnitude 
in density, leading to characteristic fragment 
masses of the order of $10^{-2}-10^{-1} \msun$. \\
Thus, it appears that a gas cloud with initial metallicity of $Z\simeq 10^{-5.1} Z_{\odot}$ and 
$f_{\rm dep} \ga 0.2$ (for comparison the Galactic value is $f_{\rm dep} = 0.47$) 
can generate low-mass fragments, subsequently leading to the formation of 
low-mass stars, consistent with the observation of HE0107-5240.
As an alternative model, a mechanism has been proposed\cite{MacKey03}  
for the formation of a population of extremely metal-poor intermediate-mass stars at very high redshifts. 
Within this model,
the formation of these stars is directly linked to the death of a very massive star exploding as a 
pair-creation supernova. Indeed, shock compression, heating and subsequent cooling to high density reduces the fragment
mass in primordial gas to $\sim 10 \msun$, allowing small mass stars to form. It might be that HE0107-5240 is the
first observed member of such a population.  \\
Our analysis shows that the discovery of HE0107-5240 does not conflict with recent studies which suggest
that the formation of low-mass stars that can survive to the present day is inhibited before a minimum level
of metal enrichment is reached. The actual value of this threshold metallicity within the currently uncertain range 
$Z_{\rm cr} = 10^{-5 \pm 1} Z_{\odot}$ (ref. 6) strongly depends on the efficiency of dust 
formation in first-generation supernova ejecta. \\
If the previous lack of very iron-deficient halo stars is to be ascribed to the 
brighter magnitude limits of previous surveys, it is likely that the Hamburg/ESO 
prism survey will lead to the identification of more stars with [Fe/H]$<-5$. 
The statistics and properties of these stars should provide 
 insights into the first epochs of cosmic star formation.
In particular, if observers were to find a large number of these objects, this might 
imply that second-generation stars were already forming with characteristic masses of 1 $\msun$, 
as in the local Universe. Conversely, if future data were to show that the bulk of metal-poor
halo stars have a metallicity in the range  $Z = 10^{-4} - 10^{-3} Z_{\odot}$ with only a
small number of $< 10^{-5} Z_{\odot}$ outliers, then this would indicate that in most cases 
small-mass stars were able to form only after this level of metallicity had been 
achieved.

\noindent
\large{\bf Acknowledgements}
We acknowledge partial support from  the Research and
Training Network ``The Physics of the Intergalactic Medium''
set up by the European Community.

\noindent
\large{\bf Competing interests statement} The authors declare that they
have no competing financial interests.

\noindent
\large{\bf Correspondence} and requests for materials should be addressed to R.S.
(e-mail: raffa@arcetri.astro.it).

\end{document}